\documentclass[oribibl]{llncs}

\usepackage{geometry}
\usepackage{graphicx}
\usepackage{amsmath}
\usepackage{amssymb}

\begin{document}

\title{Club Formation by Rational Sharing~: Content, Viability and Community Structure}

\author{W.-Y. Ng \and D.M. Chiu \and W.K. Lin}

\institute{Department of Information Engineering,\\
The Chinese University of Hong Kong\\
\email{\{wyng, dmchiu, wklin3\}@ie.cuhk.edu.hk}}

\authorrunning{W.-Y. Ng et al.}

\maketitle

\begin{abstract}
\emph{A sharing community prospers when participation and contribution are
both high.  We suggest the two, while being related decisions every peer
makes, should be given separate rational bases.  Considered as such, a basic
issue is the viability of club formation, which necessitates the modelling of
two major sources of heterogeneity, namely, peers and shared content.  This
viability perspective clearly explains why rational peers contribute (or
free-ride when they don't) and how their collective action determines
viability as well as the size of the club formed.  It also exposes another
fundamental source of limitation to club formation apart from free-riding, in
the community structure in terms of the relation between peers' interest
(demand) and sharing (supply).}
\end{abstract}

\section{Introduction}

Much current research in peer-to-peer systems focuses on performance of the
platform on which peers transact.  Even when incentives of the peers
themselves are considered, the concern is with their effects on the system
load.  Invariably, selfish peers are assumed always ready to participate.
Incentive mechanisms are then necessary to make sure they behave nicely and
do not cause excessive load. As this \emph{performance perspective} dominates
the research agenda, free-riding is often identified as a major problem, a
limiting factor to be dealt with.

In reality, free-riding \cite{ah2000} is in fact very common in open access
communities, including many successful file sharing networks on the Internet
where incentive mechanisms are often scant or absent altogether \cite{kazza,
gnutella}.  A problem in principle does not appear to be a problem in
practice. An empirical observation offers a straightforward explanation: many
peers are seemingly generous with sharing their private assets.  They upload
files for sharing, help one another in resource discovery, routing, caching,
etc.  As long as sufficiently many are contributing, free-riding may be
accommodated and the community would sustain.  But then why are they so
generous?

Our study is in part motivated by Feldman \textit{et al.}~\cite{fpcs2004} who
explains this in terms of peers' intrinsic \emph{generosity}.  Assuming that
a peer's generosity is a statistical \emph{type variable} and he would
contribute as long as his share of the total cost does not exceed it, they
show that some peers would choose to contribute \emph{even as others
free-ride}, as long as sufficiently many peers have high enough generosity.
However, how the generosity is derived is left open.  In contrast,
\cite{kstt2002} identifies a rational basis for peers to contribute based on
a utility function predicated on the benefit of information access, which is
improved as peers contribute in load sharing that eases system congestion.
They demonstrate also that a sharing community may sustain in the face of
free-riding without any explicit incentive mechanism.

\subsection{Why peers participate?}
In this paper, we contend that there remains another important basic question
to examine, namely, why peers participate in the first place.  It is of
course reasonable to assume simply that peers do so for their own benefit.
However, it is crucial to note that peers often benefit \emph{differently},
even as they participate to the same extent and contribute to the same
extent.  In a file sharing community for instance, the benefit a peer sees
would depend on whether he finds what he is interested in there, which should
vary from peer to peer.  In this regard, the studies cited above and many
others essentially assume that peers have identical interest and sees the
same benefit potential from a sharing community, which is obviously not very
realistic.

\subsection{Goods type and peer type}
As peers exhibit different interest, they would demand different things in
their participation.  It follows that the system serving a sharing community
(\emph{the club}) comprising such peers would have content with \emph{a
variety of goods}.  The availability of the goods a peer demands from the
club would determine the benefit potential he sees, and the extent of his
participation.  Two new type variables are implied: first, a type variable
for different goods, and second, a peer type variable for different interests
in the goods.  For simplicity, we may replace the second by some measure of
his \emph{propensity-to-participate}, as a proxy variable conditioned on a
given club with its particular content.

We believe the new type variables are essential to analyzing realistic
peer-to-peer systems.  We shall demonstrate this by constructing a generic
model of an \emph{information sharing club (ISC)}~\cite{nlc05_PERF_JOURNAL}
in which peers contribute by sharing information goods.  Such shared goods
then make up the club's content which in turn entice peers to participate and
contribute.

\subsection{Viability perspective}
Unlike models taking the performance perspective, the ISC model takes a
\emph{viability perspective} and focuses instead on a more primary concern of
whether the club itself has any chance to grow in size at all. In principle,
this would depend on a \emph{mutual sustenance between club membership and
content}.  This concern turns out to subsume free-riding, and reveals another
fundamental source of limitation, in \emph{community structure} in terms of
the relation between demand and supply among the sharing peers. Generally
speaking, insufficient interdependence among them for their interested goods
would limit their propensity-to-participate and the size of the club they
form.  In the worst case, the club may get stuck in a deadlock with few
members and little content, even emptiness, when there is little overlap
between their interest (demand) and potential contribution (supply).

We begin our discussion in the next section by re-visiting the question of
why rational peers contribute.  We shall derive another peer type variable --
the \emph{propensity-to-contribute} (c.f. generosity of \cite{fpcs2004}) --
from the peer's utility function of benefit and cost that arises from
\emph{both} participation and contribution.  As a result, we come up with new
insight regarding free-riding, in particular, when and why even generous
peers may cease to contribute.
Section~\ref{sec:what} introduces the concept of goods type and club content
as a distribution over goods types, and how a club may prosper on a mutual
sustenance between its membership and content.
Section~\ref{sec:isc} describes the ISC model and derives two viability
conditions. We demonstrate how the community structure affects viability in
simplistic model instances with two goods types.
In the final section, we discuss the design of incentive systems in the face
of two different sources of limitation, namely, free-riding and community
structure.

\section{\emph{Why} do rational peers share?} \label{sec:why}
\hspace*{\parindent}

Let peer $i$'s contribution (or cost) to a club be $C_i$ and the club's
benefit to him is $B_i$.  Further assume that each peer's choice of $C_i$
directly affects its benefit (therefore, $B_i$ is a function of $C_i$ among
other things). The peer's utility, a function of both the benefit and cost,
is given by $U_i(B_i, C_i)$. It is intuitive to assume that $U_i$ is
decreasing in $C_i$ and concave increasing in $B_i$. Given any particular
level of contribution $C_i$ and a corresponding level of benefit $B_i$, any
small increment of utility is given by
\[ \delta U_i(B_i, C_i) = (\partial U_i/\partial B_i) \delta B_i +
(\partial U_i/\partial C_i) \delta C_i \;.\] The (non-negative) ratio
\begin{displaymath}
-\frac{\partial U_i/\partial B_i}{\partial U_i/\partial C_i}
\end{displaymath}
then gives us the (marginal) \emph{exchange rate} of peer $i$'s contribution
to benefit.  In other words, it is the maximum amount of contribution the
peer would give in exchange for an extra unit of benefit with no net utility
loss.

Although we have not yet described how to determine the club's benefit to a
particular peer $B_i$, it suffices to say that the value of $\partial B_i/
\partial C_i$ represents the \emph{club (marginal) response} to peer i's
contribution, per current levels of benefit and cost at $(B_i, C_i)$. Of
particular interest is whether this club response to peer $i$'s
\emph{initial} contribution, viz $\partial B_i/\partial C_i|_{C_i=0}$, is
enticing enough. The answer can be different for each peer.  Specifically,
peer $i$ would contribute only if
\begin{equation}
\left(\left.\frac{\partial B_i}{\partial C_i}\right|_{C_i=0}\right)^{-1} <\;
\Gamma_i\stackrel{\triangle}{=}\left.-\frac{\partial U_i/\partial
B_i}{\partial U_i/\partial C_i}\right|_{C_i=0}\;. \label{eqn:contribute}
\end{equation}
Otherwise, he prefers not to contribute and free-ride when he actually joins
the club for the good benefit he sees.
Note that $\Gamma_i$ is a property derived from the peer's utility function,
and may serve as a type variable to characterize different peers.  We refer
to this property as a peer's \mbox{\emph{propensity-to-contribute}}.

Therefore, we have tied a peer's decision to his contribution to a club to
two quantities, namely, his propensity-to-contribute $\Gamma_i$ and the club
response $\partial B_i/\partial C_i$ which depends on the specific club
model.  This treatment of peers as rational agents is similar to the
formulation in \cite{kstt2002}. It is also compatible with \cite{fpcs2004},
in that each peer ends up being characterized by a type variable. The
difference is that \cite{fpcs2004} chose not to further explain how its type
variable of generosity, is derived. Another major difference is that
$\Gamma_i$ is dependent on $B_i$.  When $U_i$ is concave increasing in $B_i$
due to decreasing marginal return of benefit, $\Gamma_i$ would decrease as
$B_i$ increases.  In other words, \emph{improved benefit reduces a peer's
propensity-to-contribute.}

Furthermore, the club response $\partial B_i/\partial C_i$ would also tend to
decrease as $B_i$ increases. A club already offering high benefit to a peer
has less potential to reward further to incentivize his contribution.  The
club response may even reduce to naught when the maximum benefit is being
offered. In this case, peer $i$ would cease to contribute and join as a
free-rider, even when his propensity-to-contribute may not be small.

In summary, as a club prospers and peers see improving benefit, the
motivation to contribute would reduce on two causes: decrease in peer's
propensity-to-contribute \emph{and} decrease in club response.  The latter is
caused by decreasing marginal benefit of a prosperous club, which we identify
as another systematic cause of free-riding (apart from peers not being
generous enough).

\section{\emph{What} do peers share?} \label{sec:what}
\hspace*{\parindent}

Before a peer decides whether to contribute based on the marginal benefit, he
first decides whether to join based on the benefit itself, namely, $B_i$.
Research works that focus on incentive schemes often study $B_i$ as a
function of the peer's decision to contribute, namely, $C_i$, \emph{only},
whence the two decisions are not differentiated and become one. Consequently,
peers who contribute the same see the same benefit potential. This is the
assumption we call into question here.

A salient feature of many real world peer-to-peer systems is the variety of
goods being shared.  The benefit that a peer receives is dependent on what he
demands in the first place, and whether they are available in the current
club content.  Even if two peers contribute the same and demand the same, the
benefit they receive would differ in general.  Peers with similar interests
would see similar benefit potential while peers with different interests may
not.
Therefore, a peer's interest, in terms of the types of goods he demands, is
an important type variable.  However, this would be a distribution over all
goods types, which is complicated.  To account for peer $i$'s particular
interest in relation to a given club, the benefit on offer, viz. $B_i$, would
suffice.  As benefit is the primary motivation for him to participate, $B_i$
would qualify as a measure of his \emph{propensity-to-participate}, a proxy
variable for his interest conditioned on the given club.

The assumption often made, that peers contributing the same receive the same
benefit, implicitly implies a single goods type.  This would be a gross
over-simplification that ignores the variety of goods as a principal source
of peer heterogeneity.  Consequently, it would overlook important structural
properties of both club membership and content essential for a detailed
analysis of the dynamics of club formation.

\subsection{Club formation, membership and content}

A club would attract a peer by its range of shared goods, and to an extent
which depends on the availability of the goods he demands.  As a result, it
would tend to attract peers who are interested in its available content.  At
the same time, such peers contribute to the club's content with what they
share. Such is the essence of a sharing community: peers come together by
virtue of the overlap between the range of goods they share (supply) and the
range they are interested in (demand). With benefit ($B_i$) as peer's
propensity-to-participate, he determines his extent of participation.
With $\Gamma_i$ as a threshold for the club response, as his
propensity-to-contribute, he determines whether to contribute during
participation.

The mutual relation between peers' demand and supply is suggestive of
potentially complex coupled dynamics of club membership and content.  A club
would prosper on virtuous cycles of gains in membership and content, and
would decline on vicious cycles of losses in both. If and when a club
sustains would depend on the existence and size of any stable equilibrium. In
particular, when an empty club is a stable equilibrium, it signifies a
deadlock between insufficient content and insufficient membership. Otherwise,
an empty club would be unstable and self-start on the slightest perturbation,
growing towards some statistical equilibrium with a positive club size.  In
the following, we refer to such a club as being \emph{viable}, and
\emph{viability} is synonymous to instability of an empty club.

\section{A simple sharing model}\label{sec:isc}
Here we present a simple model of an information sharing club (ISC), sharing
information goods which are \emph{non-rivalrous}\footnote{Unlike rivalrous goods
such as bandwidth or storage which are congestible, non-rivalrous goods may be
consumed concurrently by many users without degradation.  Information goods are
inherently non-rivalrous as they may be readily replicated at little or no
cost.}.

The model is based on two kinds of entities: a population $\mathcal{N}$ (size
$N$) of peers, and a set $\mathcal{S}$ of information goods. In addition, we
assume the following characteristics about these peers:
\begin{enumerate}
 \item Each peer has a supply of information goods which is available
       for sharing once the peer joins the ISC.
 \item Each peer has a demand for information goods in the ISC.
\end{enumerate}

The purpose of the model is to determine, based on the characteristics of the
peer population, whether an ISC will form\footnote{Theoretically, it is also
possible that more than one clubs will form, although the analysis of
multiple club scenarios is outside the scope of this paper.}; and if so, what
is the size of this club and its content. At the heart of the model is an
assumption about how a peer decides whether he joins the club.  This decision
process can be modelled as a function of a given peer's demand function, and
a club's content. In other words, given a club with certain content (of
information goods), a peer would join the club if his demand (for information
goods) can be sufficiently met by the club.  So we complete a cycle: given
some content in the club, we can compute a peer's decision whether to join a
club; given the peers' decisions we can compute the collective content of the
club; from the content, we can compute if additional peers will join the
club.  This process can always be simulated. With suitable mathematical
abstraction, this process can also be represented as a fixed point
relationship that yields the club size (and content) as its
solution~\cite{nlc05_PERF_JOURNAL}.

In our mathematical abstraction, we let peer $i$'s demand be represented by a
probability distribution $h_i(s)$, and his supply be represented by another
distribution $g_i(s)$, where $s$ indexes the information goods in
$\mathcal{S}$. For simplicity, we normally assume $h_i(s)=h(s)$ and
$g_i(s)=g(s)$ for all $i$.  Further, a peer's decision to join a club is
based on a probability $P_i(n)$, where $n$ is the number of peers already in
the club.  Let us consider what these assumptions mean (we will come back to
what $P_i(n)$ is later).  First, by representing a peer's supply (of
information goods) using a common probability distribution, we can easily
derive the distribution of the resultant content of a club of $n$ peers.
Second, given a peer's demand and a club's content, both as probability
distributions, it is possible to characterize whether a peer joins a club as
a Bernoulli trial (where $P_i$ gives the probability of joining, and $1-P_i$
gives the probability of leaving a club). Thirdly, the composition of the
club (if formed) is not deterministic; rather it is given by the statistical
equilibrium  with peers continually joining and leaving.

As a result, the club content is given by a probability distribution composed
from the supplies of $n$ peers, where $n$ is the expected size of the club.
Each peer's joining probability, $P_i(n)$, is then given by
\begin{equation}
P_i(n) \stackrel{\triangle}{=} E_{h_i(s)}[1 - e^{-\rho_i\,\Phi(n)\, g(s)}] \label{eqn:prob}
\end{equation}
where $g(s), s\in \mathcal{S}$ is the distribution of information good shared
and available in the club over a goods type domain $\mathcal{S}$, and
$h_i(s)$ is a distribution representing of peer $i$'s interest over
$\mathcal{S}$. Here, $\Phi(n)$ represents the total quantity of information
goods found in the club if $n$ peers joined, which is given by
\begin{equation}
\Phi(n) = n \bar k + \phi_0 \label{eqn:phi}
\end{equation}
where $\bar k$ is an average peer's \emph{potential} contribution (realized
when he actually joins the club) and $\phi_0 \ge 0$ represents some
\emph{seed content} of the club. The rationale for equation (\ref{eqn:prob})
is as follows. Given any specific $s$ demanded by the considered peer, the
expected number of copies of it in the club is given by $\Phi(n) g(s)$. We
assume the probability of not finding this item in the club exponentially
diminishes with the quantity, which means the probability of finding that
item is $1 - e^{-\rho_i\,\Phi(n)\, g(s)}$. Since the demand of the considered
peer is actually a distribution $h_i(s)$, therefore this peer's likely
satisfaction is expressed as an expectation over the information goods he may
demand. Finally,
\begin{equation}
\rho_i = \rho(K_i) \in [0, 1] \label{eqn:rho}
\end{equation}
is \emph{search efficiency} that peer $i$ sees, which is made dependent
on his contribution $K_i$ by some incentive system of the club to
encourage contribution so that $\rho(K_i)$ is monotonically
increasing in $K_i$.

In this model, the benefit of the club to a peer is the extent the peer
chooses to join, viz $B_i=P_i$; the cost (of contribution), on the other hand, is simply $C_i=K_i$.

\subsection{Peer dynamics: joining and leaving}

\begin{figure}[hbtp]
\begin{center}
{\scriptsize
        \includegraphics{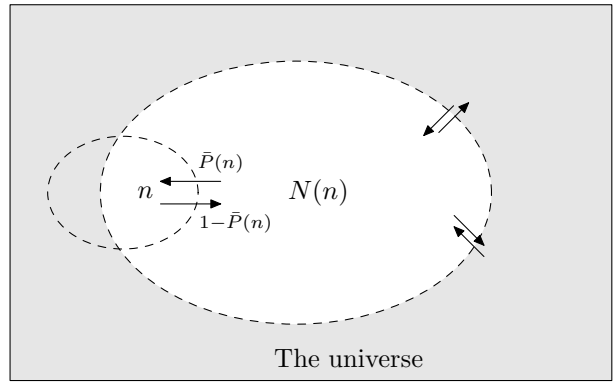}
} \caption{Peer dynamics of the ISC model} \label{figure:peer_dynamics}
\end{center}
\end{figure}

In figure (\ref{figure:peer_dynamics}) the club is depicted as the smaller
oval, and the flux of peers continually joining and leaving the club
statistically is driven by peers' propensity-to-participate (the distribution
of $P_i$). The figure also depicts the partition of the universe into the set
of potentially contributing peers (the white area), and the set of
non-contributing peers (the lightly shaded area).  This division is driven by
the population's propensity-to-contribute (the distribution of $\Gamma_i$).

By definition, non-contributing peers have no effect on $\Phi$ or $P_i$.
Therefore, we may refine the ISC model to ignore them and focus on the
potentially contributing peers (\textit{pc-peers}), namely, the population
$\mathcal{N}$ (size $N$) is the population of pc-peers and $n$ is the number
of pc-peers in the club. Subsequently, \mbox{$\bar k = 1/N \sum_{K_i > 0}
K_i$} and $K_i > 0$ is the positive contribution of pc-peer $i$.

However, the population of pc-peers is actually dependent on  $n$, viz.
$\mathcal{N}$ and $N$ should really be $\mathcal{N}(n)$ and $N(n)$. Since the
incoming rate $\bar P(n)$ \footnote{$\bar P(n) = E[P_i(n)]$ is the average
joining probability. Assuming independence between the participation and
contribution decisions, the average is the same when taken over either all
peers or the potentially contributing peers only.} depends on $n$, it gives
rise to a fixed point equation
\begin{equation}
\bar P(n) \; N(n) = n \label{eqn:equil}
\end{equation}
for the statistical equilibrium club size indicated by $n$.

As pointed out in Section~\ref{sec:why}, prosperity reduces the motivation to
contribute and some population in the white area would cease to be
(potentially) contributing and moved to the shaded part. However, we shall
assume $N(n)$ to be roughly constant here when we are studying the club's
viability property which is pre-emptive to prosperity and determined by the
club dynamics around $n=0$.
\subsection{Statistical equilibrium of membership and content}

The fixed point equation (\ref{eqn:equil}) characterizes a statistical
equilibrium club size when the rates of incoming and outgoing members of the
club are balanced~\cite{nlc05_PERF_JOURNAL}. Consequently, we have the
following proposition~:
\begin{proposition}[Sufficient viability condition]\label{propo:viable_sufficient}
\begin{equation}
 N(0) \sum_s g(s) h(s) > (\bar k \,\rho(0^+))^{-1} > 0
\end{equation}
is sufficient for the empty club to be unstable where h(s) is the expected
demand distribution\footnote{h(s) is the demand the club sees and would be a \emph{weighted}
average of peers's demand, with the weights being their demand rates.}. The
club is viable with a positive equilibrium club size that satisfies
equation (\ref{eqn:equil}).
\end{proposition}
The membership dynamics and content dynamics are closely coupled~: as
pc-peers join and leave, they alter the total shared content, inducing others
to revise their join/leave decisions. Pc-peer $i$ would contribute on joining
as long as his initial contribution could improve his utility, as discussed
in section \ref{sec:why}.

Whether some peers remain potentially contributing even when the club
is empty is essential to viability; $N(0)$ has to be strictly positive.
This leads to our second proposition~:
\begin{proposition}[Necessary viability condition]\label{propo:viable_necessary}
\begin{equation}
\rho'(0)\phi_0 > 0\;.\label{eqn:viable_necessary}
\end{equation}
\end{proposition}
This means some positive incentive to entice a peer to become a
contributor (from a non-contributor) and some
seed content, viz. $\phi_0 > 0$, are needed for a club to be viable
and not get stuck in an empty state with no contributing peers.
(However, while more seed content improves viability as well as
participation ($P_i$), it would also tend to reduce $N$ the same way that prosperity does.
The equilibrium club size would increase only if this reduction is more than offset by
the increase in $P_i$.)
%
\subsection{Community structure}
Since the basis for club formation is the overlap between peers' interest and the club's
content, an interesting question is whether peers tend to form small clubs due
to clustering of common interest, or large clubs with diverse population of peers.
Furthermore, given a club of multiple information goods, can it be decomposed
and analyzed as multiple clubs of single goods?  The answers to these questions would
shed more light on why it is important to model an information sharing club
based on content.

Suppose we have two disjoint clubs with equilibrium club sizes $n_1$ and
$n_2$, formed independently based on their respective potentially
contributing peer populations $N_1$ and $N_2$.  When they are brought
together, it is intuitive that a new club would form with at least size $n_1
+ n_2$. We refer to this as \emph{mixing} (two clubs). A key question is~:
will the size of the new club, $n$, be strictly greater than $n_1 + n_2$?

We devise two simple simulated examples below to study the effect of mixing.

First, consider a universe with only two types of information goods and two
independent clubs, each with peers interested in a different single goods
type only.  However, when they contribute, they may share some percentage $q$
of the other goods types.  The demand and supply distributions are shown in
table \ref{table:example}. Further we assume $\bar k \rho = 0.015;\; N_1 =
N_2=100$ such that both clubs are viable (when $q$ is reasonably small)
according to proposition \ref{propo:viable_sufficient}. Therefore $q$ may be
regarded as the degree of overlap between the two clubs' supply.

\begin{table}[h!]
\begin{center}
\caption{Two population and two goods type scenario}\label{table:example}
\begin{tabular}{|c||p{1.5cm}|p{1.5cm}|}
\hline
peers&
demand&
supply\tabularnewline
\hline
\hline
type $1$&
$\{1, 0\}$&
$\{1-q, q\}$\tabularnewline
\hline
type $2$&
$\{0, 1\}$&
$\{q, 1-q\}$\tabularnewline
\hline
\end{tabular}
\end{center}
\end{table}

\begin{figure}[htbp]
\begin{center}
{\scriptsize
\begin{picture}(0,0)%
\includegraphics{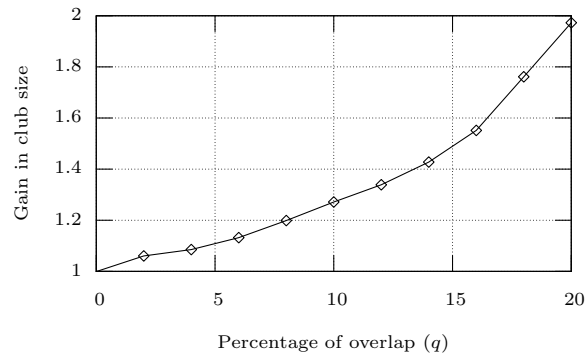}%
\end{picture}%
\begingroup
\setlength{\unitlength}{0.0200bp}%
\begin{picture}(11699,7019)(0,0)%
\put(1650,1650){\makebox(0,0)[r]{\strut{} 1}}%
\put(1650,2614){\makebox(0,0)[r]{\strut{} 1.2}}%
\put(1650,3578){\makebox(0,0)[r]{\strut{} 1.4}}%
\put(1650,4542){\makebox(0,0)[r]{\strut{} 1.6}}%
\put(1650,5506){\makebox(0,0)[r]{\strut{} 1.8}}%
\put(1650,6470){\makebox(0,0)[r]{\strut{} 2}}%
\put(1925,1100){\makebox(0,0){\strut{} 0}}%
\put(4163,1100){\makebox(0,0){\strut{} 5}}%
\put(6400,1100){\makebox(0,0){\strut{} 10}}%
\put(8638,1100){\makebox(0,0){\strut{} 15}}%
\put(10875,1100){\makebox(0,0){\strut{} 20}}%
\put(550,4060){\rotatebox{90}{\makebox(0,0){\strut{}Gain in club size}}}%
\put(6400,275){\makebox(0,0){\strut{}Percentage of overlap ($q$)}}%
\end{picture}%
\endgroup
} \caption{Effect of mixing clubs when both are viable}
\label{figure:mix_grow}
\end{center}
\end{figure}

We obtained (by simulation) the equilibrium club sizes $n_1$ and $n_2$ when
the populations $N_1$ and $N_2$ are separate, and then the equilibrium club
size $n$ when they are mixed. Figure~{(\ref{figure:mix_grow})} shows the gain
in the mixed club size, as the ratio $\frac{n}{n_1+n_2}$. With no overlap
($q=0$), the mixed club size $n$ is simply the sum of the two individual
clubs ($n_1 + n_2$).  However a larger club is formed with rapidly increasing
gain as the overlap increases. A two-fold gain results with a moderate
overlap of $20\%$.

\begin{figure}[hbtp]
\begin{center}
{\scriptsize
\begin{picture}(0,0)%
\includegraphics{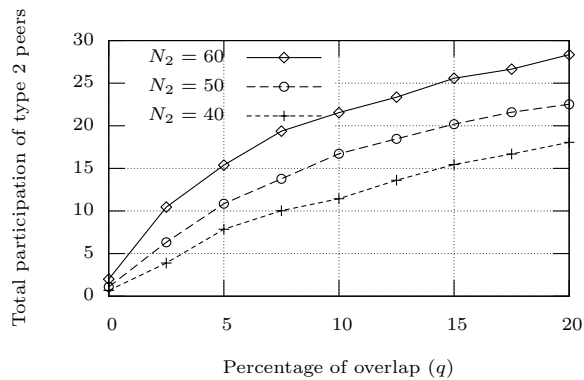}%
\end{picture}%
\begingroup
\setlength{\unitlength}{0.0200bp}%
\begin{picture}(11699,7019)(0,0)%
\put(1925,1650){\makebox(0,0)[r]{\strut{} 0}}%
\put(1925,2453){\makebox(0,0)[r]{\strut{} 5}}%
\put(1925,3257){\makebox(0,0)[r]{\strut{} 10}}%
\put(1925,4060){\makebox(0,0)[r]{\strut{} 15}}%
\put(1925,4863){\makebox(0,0)[r]{\strut{} 20}}%
\put(1925,5667){\makebox(0,0)[r]{\strut{} 25}}%
\put(1925,6470){\makebox(0,0)[r]{\strut{} 30}}%
\put(2200,1100){\makebox(0,0){\strut{} 0}}%
\put(4369,1100){\makebox(0,0){\strut{} 5}}%
\put(6538,1100){\makebox(0,0){\strut{} 10}}%
\put(8706,1100){\makebox(0,0){\strut{} 15}}%
\put(10875,1100){\makebox(0,0){\strut{} 20}}%
\put(550,4060){\rotatebox{90}{\makebox(0,0){\strut{}Total participation of type $2$ peers}}}%
\put(6537,275){\makebox(0,0){\strut{}Percentage of overlap ($q$)}}%
\put(4528,6149){\makebox(0,0)[r]{\strut{}$N_2 = 60$\hspace{.1cm}}}%
\put(4528,5599){\makebox(0,0)[r]{\strut{}$N_2 = 50$\hspace{.1cm}}}%
\put(4528,5049){\makebox(0,0)[r]{\strut{}$N_2 = 40$\hspace{.1cm}}}%
\end{picture}%
\endgroup
} \caption{Effect of mixing clubs on the non-viable club (type $2$ peers)}
\label{figure:one_bring_two}
\end{center}
\end{figure}

Our second example considers the same two populations with the total size
$N_1 + N_2 = 200$, except that $N_1 > N_2$ so that the second population do
not make a viable club this time. Figure~{(\ref{figure:one_bring_two})} shows
what happens when these two populations are mixed with various degrees of
overlap. For a good range of $N_2$, the non-viable club is able to form (as
part of the larger mixed club) with different vigor monotonically increasing
in $q$. These two examples demonstrate how the modeling of goods types help
account for club formation: a high degree of overlap between peers' supply
and demand is conducive to large club formation, and vice versa. When a large
club is formed with significant mixing, it comprises gain in membership and
content over any constituent specialized clubs, and is structurally different
from their mere union.

\subsection{Discussions}
Our reasoning and analysis has so far assumed a non-rivalrous relation among
peers.  In practice, there always are rivalrous resources which peers may
contend for sooner or later. For example, bandwidth could be scarce when
delivering large files and storage space may be limited. As a club increases
in size, \emph{diseconomy of scale} due to such contention would set in and
prohibit large club formation.  Benefit to peers would suffer, perhaps as a
consequence of a reduction in search efficiency $\rho_i$.  In this case, the
tendency to form smaller clubs with more specialized content would increase.
However, economy of scale may also be at work, most notably, due to
statistical effects (e.g. multiplexing gain) and/or network
effects\footnote{Network
 effects are diametrically opposite to sharing
costs \cite{v2003} (due to consumption of rivalrous resources, say).  They would
help make good a non-rivalrous assumption made in the presence of the
latter.},
which would increase the tendency to form larger clubs with a diverse
population of peers.

\section{Concluding remarks: incentivizing sharing}\label{sec:conclude}

The ISC example demonstrates that the overlap between peers' interest
(demand) in and sharing (supply) of the variety of goods is crucial.  This
two goods type example, while simplistic, is suggestive of the important role
played by peers who share a wider variety of goods. They may help induce
virtuous cycles that improve membership and content, resulting in a larger
club size. Further, they may help \emph{niche} peer groups, otherwise not
viable, to benefit from participating in a more mixed and larger community.
Therefore it is conceivable that such sharing creates more ``synergy" than
more specialized supplies.  In economics term, they create positive
externality and should justify positive incentives.

The viability perspective points to the importance of maintaining a large
$N(n)$ for viability is instructive to the design of incentive mechanisms. In
other words, the more potentially contributing peers the better. Contribution
should be encouraged especially when starting up a club, as viability depends
on a large enough $N(0)$.

However, it should be emphasized that encouraging contribution may not entail
discouraging free-riding.  One may imagine free-riders who are discouraged by
some negative incentive schemes simply demand less without becoming
contributing peers, and the club does not become more viable. A positive
scheme that aims to increase $N(n)$ directly is always preferred.  A
reasonable principle in economizing the use of incentive schemes would be to
focus on those peers who are bordering on free-riding, by virtue of their
propensity-to-contribute and/or the club response, to coerce them into
contributing.

In fact, a club's well-being may actually be harmed when free-riding is
overly discouraged.  First, free-riders may behave differently and become
contributors if only they stay long enough to realize more benefits in
participation.  Second, they may be useful audience to others, e.g. in
newsgroups, BBS and forums, where wider circulation may improve utility of
\emph{all} due to network effects.

However, the ISC example has made two key assumptions, namely, constant
$N(n)$ and \mbox{non-rivalrous} resources, so as to focus on the viability of
club formation.  In reality, the two limiting factors may set in at different
stages.  When the club is ``young" and/or resourceful (abundant in all
resources except those reliant on peers' sharing), viability is the critical
concern.  When it is ``grown-up" and/or contentious (in some rivalrous
resources), performance would be critical instead.  The ISC example suggests
$n$ as a key parameter to watch, which measures the club size in terms of the
\emph{total participation of potential contributing peers}. $N(n)$ would
become sensitive and drop significantly when $n$ becomes large, beyond
$n^{via}$ say.  Contention would set in as system load increases with $n$,
beyond $n^{perf}$ say.  Unless \mbox{$n^{perf} \ll n^{via}$} whence the
performance perspective always dominates, the viability perspective should
never be overlooked.

In cases where the non-rivalrous assumption is not appropriate and sharing
costs are significant \cite{v2003}, e.g. in processing, storage and/or
network bandwidth, penalizing free-riding non-contributing peers would be
more necessary to reduce their loading on the system and the contributing
peers. However, as pointed out in \cite{b_sharing_2004}, there is a trend
demonstrated strongly by sharing communities on the Internet: rivalrous
resources may become more like non-rivalrous as contention is fast reduced
due to decreasing costs and increasing excess in resources. Because of this,
it is plausible that viability would overtake performance as the central
concern in many peer-to-peer systems sooner or later, if not already so.


\appendix

\section{Appendix}
\subsection{ Proof of Proposition \ref{propo:viable_sufficient} (sufficient viability condition)}
With reference to figure~\ref{figure:peer_dynamics}, the average rate at
which pc-peers join the club of current size $n$ is:
\begin{displaymath}
r_{\mathrm{join}} = (N(n) - n)\,\bar P(n)
\end{displaymath}
while that of leaving is:
\begin{displaymath}
r_{\mathrm{leave}} = n\,(1- \bar P(n))
\end{displaymath}
Hence, the net influx of pc-peers is given by:
\begin{equation}
r_{\mathrm{influx}} \stackrel{\triangle}{=}  r_{\mathrm{join}} - r_{\mathrm{leave}}= N(n)\, \bar P(n) - n
\end{equation}
For an empty club, $n=0$,
\begin{eqnarray}
r_{\mathrm{influx}} &=& N(0)\, \bar P(0)  \nonumber\\
&=&\sum_{i\in \mathcal{N}(0)} \sum_s h_i(s) (1-e^{-\rho_i \phi_0 \,g(s)})\label{eqn:proof_4_1_1}
\end{eqnarray}

When $\phi_0 > 0$, $r_{\mathrm{influx}}$ is strictly positive since the proposition implies $N(0) > 0$. The empty club
is unstable and would grow in this case.

When $\phi_0=0$, $\bar P(0) = 0$ and $r_{\mathrm{influx}} = 0$. The empty club is at equilibrium.
However, its \emph{stability} depends on the quantity
\begin{eqnarray}
\left.\frac{\partial r_{\mathrm{influx}}}{\partial n}\right|_{n=0} &= &N(0) \bar P'(0) + \bar P(0) N'(0) - 1\nonumber \\
&=& N(0) \bar P'(0) - 1\nonumber \\
& = & \sum_{i\in \mathcal{N}(0)}\bar k \rho_i \sum_s g(s) h_i(s) - 1\nonumber \\
&\ge& N(0) \bar k \rho(0^+) - 1 \nonumber \\
&>&0
\end{eqnarray}
as implied by the proposition. The empty club is therefore also unstable and
would grow at the slightest perturbation. \hfill \textbf{Q.E.D.}
\subsection{ Proof of Proposition \ref{propo:viable_necessary} (necessary viability condition)}
According to equation (\ref{eqn:contribute}), the contribution condition of peer $i$ is given by:
\begin{displaymath}
\left(\left.\frac{\partial P_i}{\partial K_i}\right|_{K_i=0}\right)^{-1} < \Gamma_i\;.
\end{displaymath}
$N(0) > 0$ only if:
\begin{eqnarray}
\left.\frac{\partial P_i}{ \partial K_i}\right|_{n=0,\,K_i =0} &>&0 \nonumber
\end{eqnarray}
 for some peer $i$. However,
\begin{eqnarray}
\left.\frac{\partial P_i}{ \partial K_i}\right|_{n=0,\,K_i =0} = \left. \frac{\partial (\rho_i \Phi)}{\partial K_i}\sum_s g(s) h_i(s) e^{-\rho_i \,\Phi \,g(s)} \right|_{n=0,\,K_i =0} \nonumber
\end{eqnarray}
for which it is necessary that:
\begin{eqnarray}
0 < \left.\frac{\partial (\rho_i \Phi)}{ \partial K_i}\right|_{n=0,\,K_i =0} &=&\rho'(0)\phi_0 \nonumber
\end{eqnarray}
\hfill \textbf{Q.E.D.}

\bibliographystyle{ieeetran}
\bibliography{all}

\begin{thebibliography}{1}
\providecommand{\url}[1]{#1}
\csname url@rmstyle\endcsname
\providecommand{\newblock}{\relax}
\providecommand{\bibinfo}[2]{#2}
\providecommand\BIBentrySTDinterwordspacing{\spaceskip=0pt\relax}
\providecommand\BIBentryALTinterwordstretchfactor{4}
\providecommand\BIBentryALTinterwordspacing{\spaceskip=\fontdimen2\font plus
\BIBentryALTinterwordstretchfactor\fontdimen3\font minus
  \fontdimen4\font\relax}
\providecommand\BIBforeignlanguage[2]{{%
\expandafter\ifx\csname l@#1\endcsname\relax
\typeout{** WARNING: IEEEtran.bst: No hyphenation pattern has been}%
\typeout{** loaded for the language `#1'. Using the pattern for}%
\typeout{** the default language instead.}%
\else
\language=\csname l@#1\endcsname
\fi
#2}}

\bibitem{ah2000}
E.~Adar and B.~Huberman, ``Free riding on gnutella,'' \emph{First Monday},
  vol.~5, Sep 2000.

\bibitem{kazza}
\BIBentryALTinterwordspacing
``Kazaa.'' [Online]. Available: \url{http://www.kazaa.com/}
\BIBentrySTDinterwordspacing

\bibitem{gnutella}
\BIBentryALTinterwordspacing
``Gnutella.'' [Online]. Available: \url{http://www.gnutella.com/}
\BIBentrySTDinterwordspacing

\bibitem{fpcs2004}
M.~Feldman, C.~Papadimitriou, J.~Chuang, and I.~Stoica, ``Free-riding and
  whitewashing in peer-to-peer systems,'' in \emph{Proceedings of ACM SIGCOMM
  Workshop on Practice and Theory of Incentives in Networked Systems}, 2004.

\bibitem{kstt2002}
R.~Krishnan, M.~D. Smith, Z.~Tang, and R.~Telang, ``The virtual commons: Why
  free-riding can be tolerated in file sharing networks,'' in \emph{Proceedings
  of {International Conference on Information Systems}}, 2002.

\bibitem{nlc05_PERF_JOURNAL}
W.-Y. Ng, W.~K. Lin, and D.~M. Chiu, ``Statistical modelling of information
  sharing: community, membership and content,'' \emph{Performance Evaluation},
  vol. 62, issues 1-4, pp. 17--31, October, 2005.

\bibitem{v2003}
H.~R. Varian, ``The social cost of sharing,'' in \emph{Proceedings of Workshop
  on Economics of {P2P} Systems}, June 2003.

\bibitem{b_sharing_2004}
Y.~Benkler, ``{Sharing nicely}: on shareable goods and the emergence of sharing
  as a modality of economic production,'' \emph{The Yale Law Journal}, vol.
  114, pp. 273--358, 2004.

\end{thebibliography}

\end{document}